\documentstyle[12pt]{article}

\newcounter{eq}
\newcounter{sc}

\def\overleftrightarrow#1{\vbox{\ialign{##\crcr
 $\leftrightarrow$\crcr\noalign{\kern-1pt\nointerlineskip}
 $\hfil\displaystyle{#1}\hfil$\crcr}}}

\newlength{\minitwocolumn}
\begin{document}

\begin{flushright}
DPUR/TH/36\\
January, 2013\\
\end{flushright}
\vspace{20pt}

\pagestyle{empty}
\baselineskip15pt

\begin{center}
{\large\bf Classically Scale-invariant B-L Model and Dilaton Gravity
\vskip 1mm }

\vspace{20mm}
Ichiro Oda \footnote{E-mail address:\ ioda@phys.u-ryukyu.ac.jp
}

\vspace{5mm}
           Department of Physics, Faculty of Science, University of the 
           Ryukyus,\\
           Nishihara, Okinawa 903-0213, Japan.\\

\end{center}


\vspace{5mm}
\begin{abstract}
We consider a coupling of dilaton gravity to the classically scale-invariant B-L extended
standard model which has been recently proposed as a phenomenologically viable model
realizing the Coleman-Weinberg mechanism of breakdown of the electroweak symmetry.
It is shown in the present model that without recourse to the Coleman-Weinberg mechanism,
the B-L gauge symmetry is broken in the process of spontaneous symmetry breakdown 
of scale invariance at the tree level and as a result the B-L gauge field becomes massive 
via the Higgs mechanism. Since the dimensionful parameter is only the Planck mass in our model,
one is forced to pick up very small coupling constants if one wishes to realize the breaking
of the B-L symmetry at TeV scale.
\end{abstract}

\newpage
\pagestyle{plain}
\pagenumbering{arabic}


\rm
\section{Introduction}

It has been widely believed thus far that a stabilization of the scale of spontaneous symmetry
breakdown of the electroweak symmetry requires us to introduce some new physics beyond the standard
model around TeV scale. A popular scenario as such a new physics is surely supersymmetric
extensions of the standard model \cite{Martin}. In order to solve the well-known hierarchy problem,
the scale of supersymmetry breaking cannot be remote from weak scale and therefore the standard
model particles and their superpartners must have the sizable couplings around the scale.

However, the recent observations by the Large Hadron Collider (LHC) seem to exclude low energy supersymmetry
\cite{ATLAS, CMS}, so it is timely to offer the alternative idea in such a way that the scale of electroweak symmetry breaking 
can be stabilized by not TeV scale but Planck scale physics effects, and no new physics interacting
with the standard model particles at the weak scale is not needed to stabilize the weak scale.

About twenty years ago, Bardeen has proposed an interesting idea that if classical scale invariance
is imposed on the standard model, we are free from quadratic divergences and therefore can dispense with
the gauge hierarchy problem \cite{Bardeen}. In this context, let us recall that the action of the standard model has
the scale invariance except for the Higgs mass term. Since there is no negative mass squared term of the Higgs
field in the scale-invariant models, the electroweak symmetry breaking is triggered by radiative corrections 
by following the Coleman and Weinberg \cite{Coleman}. However, it is known that the Coleman-Weinberg mechanism does not
work in the standard model, so one is forced to extend the standard model with the classical scale
invariance by adding new particles. 

Since we suppose that physics at the Planck scale is directly connected with the electroweak physics,
it is natural to incorporate the gravity sector to the extensions of the standard model and ask ourselves
if the gravity sector provides a new mechanism of the electroweak symmetry breaking instead of the
Coleman-Weinberg mechanism. In this article, we wish to pursue such a possibility on the basis of dilaton
gravity \cite{Fujii}.  Incidentally, it is of interest to notice that the Higgs particle and the graviton have some
similar characteristics. Indeed, the Higgs particle is coupled universally to the mass of elementary 
particles and the graviton to the energy-momentum tensor. Furthermore, the Higgs potential in general 
generates a cosmological constant.  

Given a complex scalar field $\Phi$, we have a dimension 4 operator $\sqrt{-g} \xi \Phi^\dagger \Phi R$, 
which is renormalizable, that should be present in the effective theory. Here the coupling constant $\xi$
is obviously dimensionless and describes a non-minimal coupling between the scalar field and gravity,
so this term is invariant under global scaling transformation. As this term is not only renormalizable but
scale-invariant, we are tempted to add it as well as the kinetic term of the scalar field to some extensions of the standard model
with the classical scaling invariance \cite{Meissner, Iso}. As a concrete example in the extensions, we shall select the minimal 
B-L model which has been recently established as a phenomenologically viable model realizing the Coleman-Weinberg 
type breaking of the electroweak symmetry \cite{Iso}, but it is easy to apply our idea to any extension of the standard
model with the classical scale invariance as well.  

This article is organized as follows: In the next section, after mentioning notation and conventions,
we present the Lagrangian density of our model, derive equations of motion and discuss scale invariance. 
In Section 3, we perform a conformal transformation. In the process of spontaneous symmetry breakdown of
the scale invariance, we see that the B-L gauge field becomes massive through the Higgs mechanism. Then, with the usual
assumption of the sign of coefficients in the Higgs potential, the electroweak symmetry is spontaneously 
broken. In Section 4, we consider one-loop diagrams for the couplings between dilaton and matter fields,
and caculate the trace anomaly. The final section is devoted to discussion.

\section{Our model}

Before delving into details of our model, let us explain our notation and conventions. We mainly follow 
notation and conventions by Misner et al.'s textbook \cite{MTW}, for instance, the flat Minkowski metric
$\eta_{\mu\nu} = diag(-, +, +, +)$, the Riemann curvature tensor $R^\mu \ _{\nu\alpha\beta} = \partial_\alpha \Gamma^\mu_{\nu\beta}
- \partial_\beta \Gamma^\mu_{\nu\alpha} + \Gamma^\mu_{\sigma\alpha} \Gamma^\sigma_{\nu\beta} - \Gamma^\mu_{\sigma\beta} \Gamma^\sigma_{\nu\alpha}$, 
and the Ricci tensor $R_{\mu\nu} = R^\alpha \ _{\mu\alpha\nu}$.
The reduced Planck mass is defined as $M_p = \sqrt{\frac{c \hbar}{8 \pi G}} = 2.4 \times 10^{18} GeV$.
Through this article, we adopt the reduced Planck units where we set $c = \hbar = M_p = 1$ though we sometimes recover
the Planck mass $M_p$ for the clarification of explanation. In this units, all quantities become dimensionless. 
In order to convert a formula valid in the reduced Planck units to one valid in ordinary units,
we simply identify the non-geometrized dimension of all quantities in the equation, and then multiply
each such quantity by its appropriate conversion factor. Note that in the reduced Planck units, the
Einstein-Hilbert Lagrangian density takes the form ${\cal L}_{EH} = \frac{1}{2} \sqrt{-g} R$.

Let us start with the following Lagrangian density of our model:
\begin{eqnarray}
{\cal L} = \sqrt{-g} \left[ \xi \Phi^\dagger \Phi R - g^{\mu\nu} (D_\mu \Phi)^\dagger (D_\nu \Phi) + L_m \right],
\label{Lagr}
\end{eqnarray}
where the matter part $L_m$ is given by
\begin{eqnarray}
L_m &=& - \frac{1}{4} g^{\mu\nu} g^{\rho\sigma} F^{(1)}_{\mu\rho} F^{(1)}_{\nu\sigma} 
- \frac{1}{4} g^{\mu\nu} g^{\rho\sigma} F^{(2)}_{\mu\rho} F^{(2)}_{\nu\sigma} 
- g^{\mu\nu} (D_\mu H)^\dagger (D_\nu H) \nonumber\\
&-& \lambda_{H \Phi} (H^\dagger H)(\Phi^\dagger \Phi)
- \lambda_H (H^\dagger H)^2 - \lambda_{\Phi} (\Phi^\dagger \Phi)^2 + L_m'.
\label{Matter Lagr}
\end{eqnarray}
Here $L_m'$ denotes the remaining Lagrangian part of the standard model sector such as the Yukawa couplings and
the B-L sector such as right-handed neutrinos, which will be ignored in this article since it is irrelevant to
our argument. 

When $\Phi$ is a real scalar field, the first and second terms in ${\cal L}$ reduce to the well-known
Brans-Dicke Lagrangian density of the scalar-tensor gravity \cite{BD}
\begin{eqnarray}
{\cal L}_{BD} = \sqrt{-g} \left[ \varphi R - \omega \frac{1}{\varphi} g^{\mu\nu} 
\partial_\mu \varphi \partial_\nu \varphi \right],
\label{BD Lagr}
\end{eqnarray}
where we have defined as
\begin{eqnarray}
\varphi = \xi \Phi^2, \quad \omega = \frac{1}{8 \xi}.
\label{Omega}
\end{eqnarray}

For a generic field $\phi$, the covariant derivative $D_\mu$ is defined as \cite{Iso}
\begin{eqnarray}
D_\mu \phi &=& \partial_\mu \phi + i \left[ g_1 Q^Y A^{(1)}_\mu + (g_2 Q^Y + g_{BL} Q^{BL} ) A^{(2)}_\mu \right] \phi,
\nonumber\\
(D_\mu \phi)^\dagger &=& \partial_\mu \phi^\dagger - i \left[ g_1 Q^Y A^{(1)}_\mu + (g_2 Q^Y + g_{BL} Q^{BL} ) 
A^{(2)}_\mu \right] \phi^\dagger,
\label{Cov-der}
\end{eqnarray}
where $Q^Y$ and $Q^{BL}$ respectively denote the hypercharge and B-L charge whose corresponding gauge fields are written 
as $A^{(1)}_\mu$ and $A^{(2)}_\mu$. The charge assignment for the complex singlet scalar $\Phi$ 
and the Higgs doublet $H$ is $Q^Y(\Phi) = 0, Q^{BL}(\Phi) = 2, Q^Y(H) = \frac{1}{2}, Q^{BL}(H) = 0$. Moreover, the field strengths
for the gauge fields are defined in a usual manner as
\begin{eqnarray}
F^{(i)}_{\mu\nu} = \partial_\mu A^{(i)}_\nu - \partial_\nu A^{(i)}_\mu,
\label{F}
\end{eqnarray}
where $i = 1, 2$. Finally, let us define the potential $V(H, \Phi)$ by
\begin{eqnarray}
V(H, \Phi) = \lambda_{H \Phi} (H^\dagger H)(\Phi^\dagger \Phi)
+ \lambda_H (H^\dagger H)^2 + \lambda_{\Phi} (\Phi^\dagger \Phi)^2.
\label{V}
\end{eqnarray}

It is now worth noting that since all coupling constants in ${\cal L}$ are dimensionless, our model is manifestly invariant
under a global scale transformation. In fact, with a constant parameter $\Omega = e^\Lambda \approx 1 + \Lambda \ (|\Lambda| \ll 1)$
the scale transformation is defined as \footnote{In this article, we use the terminology that scale transformation means a global
transformation whereas conformal transformation does a local one. In some references, scale transformation is defined such that 
$g_{\mu\nu}$ keeps invariant but instead the coordinates $x^\mu$ are transformed as $x^\mu \rightarrow \tilde x^\mu = \Omega x^\mu$.
This definition might be useful in a flat Minkowski space-time \cite{Gross}.} 
\begin{eqnarray}
g_{\mu\nu} &\rightarrow& \tilde g_{\mu\nu} = \Omega^2 g_{\mu\nu},  \quad
g^{\mu\nu} \rightarrow \tilde g^{\mu\nu} = \Omega^{-2} g^{\mu\nu}, \quad
 \nonumber\\
\Phi &\rightarrow& \tilde \Phi = \Omega^{-1} \Phi, \quad
H \rightarrow \tilde H = \Omega^{-1} H,  \quad
A^{(i)}_\mu \rightarrow \tilde A^{(i)}_\mu = A^{(i)}_\mu.
\label{Scale transf}
\end{eqnarray}
Then, using the formulae $\sqrt{-g} = \Omega^{-4} \sqrt{- \tilde g}, R = \Omega^2 \tilde R$, it is straightforward to
show that ${\cal L}$ is invariant under the scale transformation (\ref{Scale transf}).
Following the Noether procedure $\Lambda J^\mu = \sum \frac{\partial {\cal L}}{\partial \partial_\mu \phi} \delta \phi$ where 
$\phi = \{g_{\mu\nu}, \Phi, \Phi^\dagger, H, H^\dagger\}$, after a little tedious calculation, the current for the scale 
transformation is obtained
\begin{eqnarray}
J^\mu = \sqrt{-g} g^{\mu\nu} \partial_\nu \left[ ( 6 \xi + 1 ) \Phi^\dagger \Phi + H^\dagger H \right].
\label{Current}
\end{eqnarray}

To prove that this current is conserved on-shell, one first needs to derive equations of motion. 
The variation of (\ref{Lagr}) with respect to the metric tensor produces Einstein's equations   
\begin{eqnarray}
2 \xi \Phi^\dagger \Phi G_{\mu\nu} = T_{\mu\nu} + T^{(\Phi)}_{\mu\nu} - 2 \xi ( g_{\mu\nu} \Box - \nabla_\mu \nabla_\nu )
(\Phi^\dagger \Phi),
\label{Einstein eq}
\end{eqnarray}
where d'Alembert operator $\Box$ is as usual defined as $\Box (\Phi^\dagger \Phi) = \frac{1}{\sqrt{-g}} \partial_\mu
(\sqrt{-g} g^{\mu\nu} \partial_\nu (\Phi^\dagger \Phi)) = g^{\mu\nu} \nabla_\mu \nabla_\nu (\Phi^\dagger \Phi)$
and the Einstein tensor is $G_{\mu\nu} = R_{\mu\nu} - \frac{1}{2} g_{\mu\nu} R$.
Here the energy-momentum tensors $T_{\mu\nu}, \ T^{(\Phi)}_{\mu\nu}$ are defined as
\begin{eqnarray}
T_{\mu\nu} &=& - \frac{2}{\sqrt{-g}} \frac{\delta(\sqrt{-g} L_m)}{\delta g^{\mu\nu}}  \nonumber\\
&=&  \sum_{i=1}^2 \left( F^{(i)}_{\mu\rho} F^{(i) \rho}_\nu - \frac{1}{4} g_{\mu\nu} F^{(i)}_{\rho\sigma} F^{(i)\rho\sigma} \right)
+ 2 (D_{(\mu} H)^\dagger (D_{\nu)} H) - g_{\mu\nu} (D_{\rho} H)^\dagger (D^{\rho} H) \nonumber\\
&-& g_{\mu\nu} V(H, \Phi),  \nonumber\\
T^{(\Phi)}_{\mu\nu} &=& - \frac{2}{\sqrt{-g}} \frac{\delta}{\delta g^{\mu\nu}} [ - \sqrt{-g} g^{\rho\sigma} (D_\rho \Phi)^\dagger 
(D_\sigma \Phi) ]  \nonumber\\
&=&  2 (D_{(\mu} \Phi)^\dagger (D_{\nu)} \Phi) - g_{\mu\nu} (D_{\rho} \Phi)^\dagger (D^{\rho} \Phi),
\label{Energy-momentum}
\end{eqnarray}
where we have used notation of symmetrization $A_{(\mu} B_{\mu)} = \frac{1}{2} (A_\mu B_\nu + A_\nu B_\mu)$. 

Next, taking the variation with respect to $\Phi^\dagger$ leads to the following equation:
\begin{eqnarray}
\xi \Phi R  + \frac{1}{\sqrt{-g}} D_\mu (\sqrt{-g} g^{\mu\nu} D_\nu \Phi) - \lambda_{H \Phi} (H^\dagger H) \Phi
- 2 \lambda_\Phi (\Phi^\dagger \Phi) \Phi = 0.
\label{Phi eq}
\end{eqnarray}
Similarly, the variation with respect to $H^\dagger$ yields the following equation of motion:
\begin{eqnarray}
\frac{1}{\sqrt{-g}} D_\mu (\sqrt{-g} g^{\mu\nu} D_\nu H) - \lambda_{H \Phi} (\Phi^\dagger \Phi) H
- 2 \lambda_H (H^\dagger H) H = 0.
\label{H eq}
\end{eqnarray}
Finally, taking the variation with respect to the gauge fields $A^{(i)}_\mu$ produces "Maxwell" equations
\begin{eqnarray}
\nabla_\rho F^{(1)\mu\rho} &=& \frac{1}{2} i g_1 \left[ H^\dagger (D^\mu H) - H (D^\mu H)^\dagger \right], \nonumber\\
\nabla_\rho F^{(2)\mu\rho} &=& \frac{1}{2} i g_2 \left[ H^\dagger (D^\mu H) - H (D^\mu H)^\dagger \right]
+ 2 i g_{BL} \left[ \Phi^\dagger (D^\mu \Phi) - \Phi (D^\mu \Phi)^\dagger \right].
\label{Maxwell eq}
\end{eqnarray}

Now we wish to prove that the current (\ref{Current}) for the scale transformation is indeed conserved on-shell
by using these equations of motion.
Taking the divergence of the current, we have  
\begin{eqnarray}
\partial_\mu J^\mu = \sqrt{-g} \Box \left[ ( 6 \xi + 1 ) \Phi^\dagger \Phi + H^\dagger H \right].
\label{Div-Current}
\end{eqnarray}
In order to show that the expression in the RHS vanishes on-shell, let us first take the trace
of Einstein's equations (\ref{Einstein eq}) whose result is given by
\begin{eqnarray}
\xi \Phi^\dagger \Phi R &=& (D_\mu H)^\dagger (D^\mu H) + (D_\mu \Phi)^\dagger (D^\mu \Phi)
+ 3 \xi \Box (\Phi^\dagger \Phi) \nonumber\\
&+& 2 \left[ \lambda_{H \Phi} (H^\dagger H) (\Phi^\dagger \Phi) + \lambda_H (H^\dagger H)^2
+ \lambda_\Phi (\Phi^\dagger \Phi)^2 \right].
\label{Trace-Einstein eq}
\end{eqnarray}
Next, multiplying Eq. (\ref{Phi eq}) by $\Phi^\dagger$, and then eliminating the term involving
the scalar curvature, i.e. $\xi \Phi^\dagger \Phi R$, with the help of Eq. (\ref{Trace-Einstein eq}), we obtain
\begin{eqnarray}
&{}& (D_\mu H)^\dagger (D^\mu H) + (D_\mu \Phi)^\dagger (D^\mu \Phi)
+ 3 \xi \Box (\Phi^\dagger \Phi)  \nonumber\\
&+& \frac{1}{\sqrt{-g}} \left[ \Phi^\dagger D_\mu (\sqrt{-g} g^{\mu\nu} D_\nu \Phi) 
+ H^\dagger D_\mu (\sqrt{-g} g^{\mu\nu} D_\nu H) \right]   
= 0.
\label{Combined eq}
\end{eqnarray}

At this stage, it is useful to introduce a generalized covariant derivative defined
as ${\cal D}_\mu = D_\mu + \Gamma_\mu$, and using this derivative Eq. (\ref{Combined eq})
can be rewritten as
\begin{eqnarray}
\Phi^\dagger {\cal D}_\mu {\cal D}^\mu \Phi + ({\cal D}_\mu \Phi)^\dagger ({\cal D}^\mu \Phi)
+ 3 \xi {\cal D}_\mu {\cal D}^\mu (\Phi^\dagger \Phi) + H^\dagger {\cal D}_\mu {\cal D}^\mu H
+ ({\cal D}_\mu H)^\dagger ({\cal D}^\mu H) = 0.
\label{Re-Combined eq}
\end{eqnarray}
Then, adding its Hermitian conjugation to Eq. (\ref{Re-Combined eq}), we arrive at 
\begin{eqnarray}
{\cal D}_\mu {\cal D}^\mu \left[ (6 \xi + 1) \Phi^\dagger \Phi + H^\dagger H \right] = 0.
\label{Re-Combined eq2}
\end{eqnarray}
The quantity in the square bracket is a scalar and neutral under two U(1) charges, 
we obtain 
\begin{eqnarray}
\Box \left[ (6 \xi + 1) \Phi^\dagger \Phi + H^\dagger H \right] = 0.
\label{Re-Combined eq3}
\end{eqnarray}
Using this equation, the RHS in Eq. (\ref{Div-Current}) is certainly vanishing, by which
we can prove that the current of the scale transformation is conserved on-shell.

\section{Conformal transformation and spontaneous symmetry breakdown of scale invariance}

Now we are ready to discuss spontaneous symmetry breakdown of scale invariance in our
model. In ordinary examples of spontaneous symmetry breakdown in the framework of
quantum field theories, one is accustomed to dealing with a potential which has the shape of 
the Mexican hat type and therefore induces the symmetry breaking in a natural way, but
the same recipe cannot be applied to general relativity because of a lack of such a
potential. \footnote{In the case of massive gravity, a similar situation occurs in breaking
the general coordinate invariance spontaneously \cite{Oda}.}

However, a very interesting recipe which induces spontaneous symmetry breakdown of scale
invariance via conformal transformation has been known \cite{Fujii}. Recall that we have started with 
a scale-invariant theory with only dimensionless coupling constants. But in the process
of conformal transformation, one cannot refrain from introducing the quantity with mass
dimension, which is the Planck mass $M_p$ in the present context, to match the dimensions 
of an equation and consequently scale invariance is spontaneously broken. Of course,
the absence of a potential which induces symmetry breaking makes it impossible to
investigate a stability of the selected solution, but the very existence of the solution
including the Planck mass with mass dimension justifies the claim that this phenomenon is
nothing but a sort of spontaneous symmetry breakdown. Note that a similar phenomenon can
be also seen in spontaneous compactification in the Kaluza-Klein theories.

The first step towards obtaining spontaneous symmetry breakdown of scale invariance is to 
find a suitable conformal transformation which transforms dilaton
gravity in the Jordan frame to general relativity with matters in the Einstein frame.
It is then convenient to parametrize the complex scalar field $\Phi$ in terms of two real
fields, those are $\Omega$ (or $\sigma$) and $\theta$ in polar form as
\begin{eqnarray}
\Phi(x) = \frac{1}{\sqrt{2 \xi}} \Omega(x) e^{i \alpha \theta(x)} 
=  \frac{1}{\sqrt{2 \xi}} e^{\zeta \sigma(x) + i \alpha \theta(x)},
\label{Parametrization}
\end{eqnarray}
where $\Omega(x) = e^{\zeta \sigma(x)}$ is a local parameter field in contrast with a global
parameter in the scale transformation (\ref{Scale transf}). The constants $\zeta, \alpha$ will be 
determined shortly.

Let us consider the following conformal transformation:
\begin{eqnarray}
g_{\mu\nu} &\rightarrow& \tilde g_{\mu\nu} = \Omega^2(x) g_{\mu\nu},  \quad
g^{\mu\nu} \rightarrow \tilde g^{\mu\nu} = \Omega^{-2}(x) g^{\mu\nu}, \quad
\nonumber\\
H &\rightarrow& \tilde H = \Omega^{-1}(x) H,  \quad
A^{(i)}_\mu \rightarrow \tilde A^{(i)}_\mu = A^{(i)}_\mu.
\label{Conformal transf}
\end{eqnarray}
Note that apart from the local property of $\Omega(x)$, this conformal transformation is
different from the scale transformation (\ref{Scale transf}) in that the complex scalar
field $\Phi$ is not transformed at all. Under the conformal transformation (\ref{Conformal transf}),
the scalar curvature is transformed as
\begin{eqnarray}
R = \Omega^2 ( \tilde R + 6 \tilde \Box f - 6 \tilde g^{\mu\nu} \partial_\mu f \partial_\nu f ),
\label{Curvature}
\end{eqnarray}
where we have defined as $f = \log \Omega = \zeta \sigma$ and $\tilde \Box f = \frac{1}{\sqrt{- \tilde g}} 
\partial_\mu (\sqrt{- \tilde g} \tilde g^{\mu\nu} \partial_\nu f) = \tilde g^{\mu\nu} 
\tilde \nabla_\mu \tilde \nabla_\nu f$.

With the critical choice
\begin{eqnarray}
\xi \Phi^\dagger \Phi = \frac{1}{2} \Omega^2 = \frac{1}{2} e^{2 \zeta \sigma},
\label{Choice}
\end{eqnarray}
the first term in (\ref{Lagr}) reads the Einstein-Hilbert term (plus the kinetic term of the scalar field $\sigma$)
up to a surface term as follows:
\begin{eqnarray}
\sqrt{-g} \xi \Phi^\dagger \Phi R &=& \Omega^{-4} \sqrt{- \tilde g} \frac{1}{2} \Omega^2
\Omega^2 ( \tilde R + 6 \tilde \Box f - 6 \tilde g^{\mu\nu} \partial_\mu f \partial_\nu f )  \nonumber\\
&=& \sqrt{- \tilde g} \left( \frac{1}{2} \tilde R - 3 \zeta^2 \tilde g^{\mu\nu} \partial_\mu \sigma \partial_\nu \sigma \right).
\label{1st term}
\end{eqnarray}
Then, the second term in (\ref{Lagr}) is cast to the form
\begin{eqnarray}
- \sqrt{-g} g^{\mu\nu} (D_\mu \Phi)^\dagger (D_\nu \Phi) 
= - \frac{1}{2 \xi} \sqrt{- \tilde g} \tilde g^{\mu\nu} \left( \zeta^2
\partial_\mu \sigma \partial_\nu \sigma + 4 g^2_{BL} M_p^2 B_\mu^{(2)} B_\nu^{(2)} \right),
\label{2nd term}
\end{eqnarray}
where we have choosen $\alpha = 2 g_{BL}$ for convenience, recovered the Planck mass $M_p$ for the clarification,
and defined a new massive gauge field $B_\mu^{(2)}$ as
\begin{eqnarray}
B_\mu^{(2)} = A_\mu^{(2)} + \partial_\mu \theta.
\label{B-field}
\end{eqnarray}

It is worthwhile to stress that in the process of conformal transformation we have had to introduce
the mass scale into a theory having no dimensional constants, thereby inducing the breaking of the scale
invariance. More concretely, to match the dimensions in the both sides of the equation
the Planck mass $M_p$ must be introduced in the ciritical choice (\ref{Choice}) 
(recovering the Planck mass)
\begin{eqnarray}
\xi \Phi^\dagger \Phi = \frac{1}{2} \Omega^2 M_p^2 = \frac{1}{2} e^{2 \zeta \sigma} M_p^2.
\label{Choice2}
\end{eqnarray}
It is also remarkable to notice that in the process of spontaneous symmetry breakdown of the scale invariance 
the Nambu-Goldstone boson $\theta$ is absorbed into the gauge field $A_\mu^{(2)}$ corresponding to
the B-L U(1) symmetry as a longitudinal mode and as a result $B_\mu^{(2)}$ acquires a mass, 
which is nothing but the Higgs mechanism! In other words,
the B-L symmetry is broken at the same time and the same energy scale that the scale symmetry is spontaneously broken.
The size of the mass $M_B$ of $B_\mu^{(2)}$ can be read off from (\ref{2nd term}) as $M_B = \frac{2}{\sqrt{\xi}} g_{BL} M_p$
which is also equal to the energy scale that the scale invariance is broken. Note that this energy scale depends on
the two unknown parameters $\xi, g_{BL}$ in the theory at hand.

Adding (\ref{1st term}) and (\ref{2nd term}) and defining $\zeta^{-2} = 6 + \frac{1}{\xi}$ (by which
the kinetic term for the $\sigma$ field becomes a canonical form), one has
\begin{eqnarray}
\sqrt{-g} \left[ \xi \Phi^\dagger \Phi R - g^{\mu\nu} (D_\mu \Phi)^\dagger (D_\nu \Phi) \right] 
= \sqrt{- \tilde g} \left( \frac{1}{2} \tilde R - \frac{1}{2} \tilde g^{\mu\nu} \partial_\mu \sigma \partial_\nu \sigma
- \frac{2}{\xi} g^2_{BL} M_p^2 B_\mu^{(2)} B^{(2)\mu} \right).
\label{1st+2nd term}
\end{eqnarray}
Note again that the first term coincides with the Einstein-Hilbert term in general relativity. To put differently,
via conformal transformation we have moved from the Jordan frame to the Einstein frame.

In a similar way, the Lagrangian density of matter fields can be written in the Einstein frame as
\begin{eqnarray}
{\cal L}_m &\equiv& \sqrt{-g} L_m \nonumber\\
&=& \sqrt{- \tilde g} \left[ - \frac{1}{4} \sum_{i=1}^2 \tilde g^{\mu\nu} \tilde g^{\rho\sigma} 
\tilde F^{(i)}_{\mu\rho} \tilde F^{(i)}_{\nu\sigma} 
- \tilde g^{\mu\nu} (\tilde D_\mu \tilde H)^\dagger (\tilde D_\nu \tilde H)
- V(\tilde H) \right].
\label{Matter Lagr in Ein}
\end{eqnarray}
Here the field strengths $\tilde F^{(i)}_{\mu\nu}$, the covariant derivative $\tilde D_\mu$
and the potential term $V(\tilde H)$ (for which the Planck mass is written explicitly) are defined as
\begin{eqnarray}
\tilde F^{(1)}_{\mu\nu} &=& \partial_\mu \tilde A^{(1)}_\nu - \partial_\nu \tilde A^{(1)}_\mu, \nonumber\\
\tilde F^{(2)}_{\mu\nu} &=& \partial_\mu B^{(2)}_\nu - \partial_\nu B^{(2)}_\mu, \nonumber\\
\tilde D_\mu &=& D_\mu + \zeta (\partial_\mu \sigma), \nonumber\\
V(\tilde H) &=& \frac{1}{4 \xi^2} \lambda_\Phi M_p^4 + \frac{1}{2 \xi} \lambda_{H \Phi} M_p^2 (\tilde H^\dagger \tilde H)
+ \lambda_H (\tilde H^\dagger \tilde H)^2.
\label{Def}
\end{eqnarray}

For spontaneous symmetry breakdown of the electroweak symmetry, let us assume
\begin{eqnarray}
\lambda_{H \Phi} < 0, \quad \lambda_H > 0.
\label{EW SSB}
\end{eqnarray}
Then, parametrizing $\tilde H^T = (0, v + \tilde h) e^{i \varphi}$, up to a cosmological constant 
the potential is reduced to the form
\begin{eqnarray}
V(\tilde H) = \frac{1}{2} m_h^2 \tilde h^2 + \sqrt{2 \lambda_H} m_h \tilde h^3 + \lambda_H \tilde h^4, 
\label{Pot}
\end{eqnarray}
where we have defined as
\begin{eqnarray}
v^2 = \frac{1}{4 \xi} \frac{|\lambda_{H \Phi}|}{\lambda_H} M_p^2 = \frac{m_h^2}{8 \lambda_H}, \quad
m_h^2 = \frac{2}{\xi} |\lambda_{H \Phi}| M_p^2.
\label{Parameters}
\end{eqnarray}
By the order estimate, $m_h \approx v \approx 10^{-16} M_p$, which requires us 
to take two conditions
\begin{eqnarray}
\lambda_H \approx 1, \quad \frac{|\lambda_{H \Phi}|}{\xi} \approx 10^{-32}.
\label{Condition}
\end{eqnarray}
The former condition is a desired condition which means that the Higgs self-coupling 
is strong and in the regime of the order 1 at the low energy. On the 
other hand, the latter condition is an original one in the present
theory.  In our model, there are four unknown parameters $\xi, \tilde g, g_{BL}$ and 
$\lambda_{\Phi}$. If we assume $\xi \approx 1$, the latter condition implies $|\lambda_{H \Phi}|
\approx 10^{-32}$ which is very small compared to the value $|\lambda_{H \Phi}|
\approx 10^{-3}$ which was derived by using the Coleman-Weinberg
mechanism in Ref. \cite{Iso}. But at present there is no experimental constraints on the value of $\xi$
coming from gravity sector, so one cannot understand the relation between our theory and the theory in Ref. \cite{Iso}. 
Moreover, as mentioned above, 
the scale of the B-L symmetry breaking is approximately given by the mass of $B^{(2)}_\mu$, 
which is $M_B = \frac{2}{\sqrt{\xi}} g_{BL} M_p$. Since this expression is also dependent on
the unknown parameters $\xi$ and $g_{BL}$, one cannot predict a precise value of the scale
of the B-L symmetry breaking either.

Finally, let us comment on the physical meaning of a scalar field $\sigma$, which we call
"dilaton". The dilaton is a massless particle and interact with the other fields only
through the covariant derivative $\tilde D_\mu = D_\mu + \zeta (\partial_\mu \sigma)$,
but owing to its nature of the derivative coupling, at the low energy this coupling is 
so small that it is difficult to detect the dilaton experimentally. 

To clarify the physical meaning more closely, it is useful to evaluate the dilatation current 
$J^\mu$ in (\ref{Current}) in the Einstein frame. The result reads
\begin{eqnarray}
J^\mu = \sqrt{- \tilde g} \tilde g^{\mu\nu} \left[ \zeta^{-1} \partial_\nu \sigma
+ (\partial_\nu + 2 \zeta \partial_\nu \sigma) \tilde H^\dagger \tilde H \right].
\label{Current2}
\end{eqnarray}
The corresponding charge is defined as $Q_D = \int d^3 x J^0$. But this charge
does not annihilate the vacuum because of the first term which is linear in $\sigma$
\begin{eqnarray}
Q_D | 0 > \neq 0.
\label{Vacuum}
\end{eqnarray}
Of course, it is also possible to show $\partial_\mu J^\mu = 0$  in terms of equations of motion in the Einstein frame
as proved in the Jordan frame before.  It therefore turns out that the dilaton $\sigma$ plays a role of the Nambu-Goldstone
boson associated with spontaneous symmetry breakdown of the scale invariance.

\section{One-loop effects}

In this section, we would like to depart from the classical analysis and move on to the evaluation of
one-loop diagrams for the coupling between dilaton and matter fields. As will be seen later, our calculation will lead
to trace anomaly in the model at hand. Note that we are not ambitious enough to quantize the metric
tensor field, but consider only radiative corrections between dilaton and matter fields in the weak field approximation.
One of motivations behind this study is to show that the conditions in Eq. (\ref{Condition}), which are important for
realizing our mechanism of electroweak symmetry breakdown at the tree level, are related to
the trace anomaly so they do not change so much even in the one-loop level as long as the violation of scale invariance is "mild".

As a regularization method, we make use of the method of continuous space-time dimensions for which
we rewrite previous results in arbitrary $D$ dimensions \cite{Fujii2}. Like the dimensional regularization,
the diveregences will appear as poles $\frac{1}{D-4}$ at the one-loop level, which are cancelled by
the factor $D-4$ that multiplies the dilaton coupling, thereby giving us a finite result yielding an effective
interaction term. 

In general $D$ space-time dimensions, as a generalization of Eq.  (\ref{Conformal transf}),
the conformal transformation is defined as
\begin{eqnarray}
\hat g_{\mu\nu} &=& \Omega^2(x) g_{\mu\nu},  \quad \hat g^{\mu\nu} = \Omega^{-2}(x) g^{\mu\nu}, \quad
\hat \Phi = \Omega^{- \frac{D-2}{2}} \Phi, \nonumber\\
\hat H &=& \Omega^{- \frac{D-2}{2}} H,  \quad \hat A^{(i)}_\mu = \Omega^{- \frac{D-4}{2}} A^{(i)}_\mu.
\label{Conformal transf 2}
\end{eqnarray}
Although, under the conformal transformation (\ref{Conformal transf 2}),
the scalar curvature is transformed as
\begin{eqnarray}
R = \Omega^2 \left[ \hat R + 2 (D-1) \hat \Box f - (D-1) (D-2) \hat g^{\mu\nu} \partial_\mu f \partial_\nu f \right],
\label{Curvature 2}
\end{eqnarray}
we set $D=4$ in this expression since we do not quantize the metric tensor and therefore do not have poles 
from the curvature. 

The critical choice (\ref{Choice}) is changed to be 
\begin{eqnarray}
\xi \Phi^\dagger \Phi = \frac{1}{2} \Omega^{D-2}, \quad \Omega = e^{\frac{2}{D-2} \zeta \sigma}.
\label{Choice 2}
\end{eqnarray}
With this choice (\ref{Choice 2}) and the conformal transformation (\ref{Conformal transf 2}),
the first term in (\ref{Lagr}) takes the similar form to (\ref{1st term})  
\begin{eqnarray}
\sqrt{-g} \xi \Phi^\dagger \Phi R = 
\sqrt{- \hat g} \left( \frac{1}{2} \hat R - 3 \zeta^2 \hat g^{\mu\nu} \partial_\mu \sigma \partial_\nu \sigma \right).
\label{1st term 2}
\end{eqnarray}
Similarly, the second term in (\ref{Lagr}) is changed to the form
\begin{eqnarray}
- \sqrt{-g} g^{\mu\nu} (D_\mu \Phi)^\dagger (D_\nu \Phi) 
= - \frac{1}{2 \xi} \sqrt{- \hat g} \hat g^{\mu\nu} \left( \zeta^2
\partial_\mu \sigma \partial_\nu \sigma + 4 \hat g^2_{BL} M_p^2 \hat B_\mu^{(2)} \hat B_\nu^{(2)} \right),
\label{2nd term 2}
\end{eqnarray}
where as before we have choosen $\alpha = 2 g_{BL}$, but we have introduced new definitions
\begin{eqnarray}
\hat g_{BL} = \Omega^{\frac{D}{2}-2} g_{BL}, \quad \hat B_\mu^{(2)} = \hat A_\mu^{(2)} + \partial_\mu \theta.
\label{B-field 2}
\end{eqnarray}

Putting (\ref{1st term 2}) and (\ref{2nd term 2}) together yields a similar expression to  (\ref{1st+2nd term})
\begin{eqnarray}
\sqrt{-g} \left[ \xi \Phi^\dagger \Phi R - g^{\mu\nu} (D_\mu \Phi)^\dagger (D_\nu \Phi) \right] 
= \sqrt{- \hat g} \left( \frac{1}{2} \hat R - \frac{1}{2} \hat g^{\mu\nu} \partial_\mu \sigma \partial_\nu \sigma
- \frac{2}{\xi} \hat g^2_{BL} M_p^2 \hat B_\mu^{(2)} \hat B^{(2)\mu} \right).
\label{1st+2nd term 2}
\end{eqnarray}

On the other hand, the Lagrangian density of matter fields turns out to depend on the dilaton field $\sigma$
in a non-trivial manner in general $D$ space-time dimensions.  The result is given by
\begin{eqnarray}
{\cal L}_m &\equiv& \sqrt{-g} L_m \nonumber\\
&=& \sqrt{- \hat g} \left[ - \frac{1}{4} \sum_{i=1}^2 \hat g^{\mu\nu} \hat g^{\rho\sigma} 
\hat F^{(i)}_{\mu\rho} \hat F^{(i)}_{\nu\sigma} 
- \hat g^{\mu\nu} (\hat D_\mu \hat H)^\dagger (\hat D_\nu \hat H)
- V(\hat H) \right],
\label{Matter Lagr in Ein 2}
\end{eqnarray}
where various quantities are defined as 
\begin{eqnarray}
\hat F^{(1)}_{\mu\nu} &=& \Omega^{2- \frac{D}{2}} F^{(1)}_{\mu\nu} 
= \partial_\mu \hat A^{(1)}_\nu + \frac{D-4}{D-2} \zeta \partial_\mu \sigma \hat A^{(1)}_\nu
- (\mu \leftrightarrow \nu), \nonumber\\
\hat F^{(2)}_{\mu\nu} &=& \Omega^{2- \frac{D}{2}} F^{(2)}_{\mu\nu} 
= \partial_\mu \hat B^{(2)}_\nu + \frac{D-4}{D-2} \zeta \partial_\mu \sigma \hat B^{(2)}_\nu
- (\mu \leftrightarrow \nu), \nonumber\\
\hat D_\mu \hat H &=& \left[ \partial_\mu + \frac{i}{2} (\hat g_1 \hat A^{(1)}_\mu + \hat g_2 \hat A^{(2)}_\mu)
+ \zeta (\partial_\mu \sigma) \right] \hat H, \nonumber\\
\hat g_{(i)} &=& \Omega^{\frac{D}{2}-2} g_{(i)}, \nonumber\\
V(\hat H) &=& e^{\frac{2(D-4)}{D-2} \zeta \sigma} \left[ \frac{1}{4 \xi^2} \lambda_\Phi M_p^4 
+ \frac{1}{2 \xi} \lambda_{H \Phi} M_p^2 (\hat H^\dagger \hat H)
+ \lambda_H (\hat H^\dagger \hat H)^2 \right].
\label{Def 2}
\end{eqnarray}

Now we wish to consider couplings between the dilaton field $\sigma$ and matter fields which
vanish at the classical level ($D=4$) but provide a finite contribution interpreted as the
trace anomaly. For simplicity of presentation, let us first switch off the U(1) fields and focus on the coupling between the
dilaton field and the Higgs field. After that, the coupling between the dilaton field and the U(1) fields
will be considered. In the weak field approximation, let us extract terms linear in the dilaton $\sigma$ in 
$V(\hat H)$ as
\begin{eqnarray}
e^{\frac{2(D-4)}{D-2} \zeta \sigma} \approx 1 + (D-4) \zeta \sigma.
\label{EXP}
\end{eqnarray}
Then, with the SSB ansatz  (\ref{EW SSB}) and the parametrization $\hat H^T = (0, v + \hat h) e^{i \varphi}$
the potential $V(\hat H)$ can be expanded as 
\begin{eqnarray}
V(\hat H) = V^{(0)} (\hat H) + V^{(1)} (\hat H), 
\label{Pot 2}
\end{eqnarray}
where we have defined as
\begin{eqnarray}
V^{(0)} (\hat H) &=& \frac{1}{2} m_h^2 \hat h^2 + \sqrt{2 \lambda_H} m_h \hat h^3 + \lambda_H \hat h^4,
\nonumber\\
V^{(1)} (\hat H) &=& (D-4) \zeta V^{(0)} (\hat H) \sigma.
\label{Pot 3}
\end{eqnarray}

Here we want to consider three-point (external particles are 2 Higgs $\hat h$ and 1 dilaton $\sigma$),
one-loop diagrams.
Inspection of the vertices reveals that we have three types of one-loop divergent diagrams in which
the Higgs field is circulating and one dilaton field, whose momentum is vanishing, couples. Note that 
the divergences stemming from the Higgs one-loop diagrams provide us with poles $\frac{1}{D-4}$
which cancel the factor $D-4$ multiplying the dilaton coupling in $V^{(1)} (\hat H)$, thereby yielding
a finite contribution. 

One type of one-loop divergent diagram, which we call the diagram (A), is given by the Higgs loop 
to which the dilaton couples by the vertex $- (D-4) 4! \zeta \lambda_H$ in $V^{(1)} (\hat H)$. 
The corresponding amplitude ${\cal{T}}_A$ is of form 
\begin{eqnarray}
{\cal{T}}_A &=& - i (D-4) 4! \zeta \lambda_H  \int \frac{d^D k}{(2 \pi)^D} \frac{1}{k^2 + m_h^2}
\nonumber\\
&=& \frac{3}{\pi^2} \zeta  \lambda_H  m_h^2,
\label{T_A}
\end{eqnarray}
where we have used  the familiar formula 
\begin{eqnarray}
\int \frac{d^D k}{(2 \pi)^D} \frac{1}{k^2 + m_h^2}
= \frac{i \pi^2}{(2 \pi)^4} (m_h^2)^{\frac{D}{2} -1} \Gamma(1- \frac{D}{2}),
\label{I_A}
\end{eqnarray}
and the property of the gamma function $\Gamma(m+1) = m \Gamma(m)$.

The second type of one-loop divergent diagram, which we call the diagram (B), is given by the Higgs loop 
to which the dilaton couples by the vertex $- (D-4) \zeta m_h^2$ in $V^{(1)} (\hat H)$ and
with the Higgs self-coupling vertex $- 4! \lambda_H$ in $V^{(0)} (\hat H)$. 
The amplitude ${\cal{T}}_B$ is calculated as 
\begin{eqnarray}
{\cal{T}}_B &=& i (D-4) 4! \zeta \lambda_H m_h^2 \int \frac{d^D k}{(2 \pi)^D} \frac{1}{(k^2 + m_h^2)^2}
\nonumber\\
&=& \frac{3}{\pi^2} \zeta  \lambda_H  m_h^2,
\label{T_B}
\end{eqnarray}
where we have used  the equation
\begin{eqnarray}
\int \frac{d^D k}{(2 \pi)^D} \frac{1}{(k^2 + m_h^2)^2}
= - \frac{\partial}{\partial m_h^2} \int \frac{d^D k}{(2 \pi)^D} \frac{1}{k^2 + m_h^2}
= \frac{i}{16\pi^2} \Gamma(2- \frac{D}{2}).
\label{I_B}
\end{eqnarray}

The final type of one-loop diagram, which we call the diagram (C), is a little more involved and given by the Higgs loop 
to which the dilaton couples by the vertex $- (D-4) 3! \zeta \sqrt{2 \lambda_H} m_h$ in $V^{(1)} (\hat H)$ and
with the Higgs self-coupling vertex $- 3! \sqrt{2 \lambda_H} m_h$ in $V^{(0)} (\hat H)$.  
The amplitude ${\cal{T}}_C$ reads
\begin{eqnarray}
{\cal{T}}_C &=& 2 i (D-4) \zeta \left(- 3! \sqrt{2 \lambda_H} m_h \right)^2  \int \frac{d^D k}{(2 \pi)^D} 
\frac{1}{(k^2 + m_h^2)\left[(q-k)^2 + m_h^2 \right]}
\nonumber\\
&=& \frac{18}{\pi^2} \zeta  \lambda_H  m_h^2,
\label{T_C}
\end{eqnarray}
where $q$ is the external momentum of the Higgs field. In order to reach the final result in Eq.  (\ref{T_C}), 
we have evaluated the integral as follows: 
\begin{eqnarray}
I &=& \int d^D k \frac{1}{(k^2 + m_h^2)\left[(q-k)^2 + m_h^2 \right]}
\nonumber\\
&=& \int_0^1 d x \int d^D k \frac{1}{\left[ (k^2 + m_h^2) (1-x) + ( (q-k)^2 + m_h^2) x \right]^2}
\nonumber\\
&=& \int_0^1 d x \int d^D k \frac{1}{\left[ (k- xq)^2 + m_h^2 + x (1-x) q^2 \right]^2}
\nonumber\\
&=& \int_0^1 d x \int d^D k \frac{1}{\left[ k^2 + m_h^2 + x (1-x) q^2 \right]^2}
\nonumber\\
&=& \int_0^1 d x \ i  \pi^2 \Gamma(2-\frac{D}{2}) (m_h^2)^{\frac{D}{2} -2} (1 - x + x^2)^{\frac{D}{2} -2}
\nonumber\\
&=& i \pi^2 \Gamma(2-\frac{D}{2}).
\label{Integral-C}
\end{eqnarray}
Here at the second equality, we have used the Feynman parameter formula 
\begin{eqnarray}
\frac{1}{ab} = \int_0^1 d x \frac{1}{\left[ a x + b  (1-x) \right]^2},
\label{Feynman}
\end{eqnarray}
and at the fourth equality, we have shifted the momentum $k - x q \rightarrow k$, which is allowed
since the integral is now finite owing to the regularization.

Thus, adding three types of contributions, we have 
\begin{eqnarray}
{\cal{T}} = {\cal{T}}_A + {\cal{T}}_B +  {\cal{T}}_C = \frac{24}{\pi^2} \zeta  \lambda_H  m_h^2,
\label{All-T}
\end{eqnarray}
At this stage, it is straightforward to derive the following relation
\begin{eqnarray}
\frac{1}{\sqrt{- \hat g}} < \partial_\mu J^\mu > = \frac{1}{\zeta} (D-4) < V^{(0)} (\hat H) >
= \frac{24}{\pi^2} \lambda_H  m_h^2 M_p^2.
\label{Trace anomaly}
\end{eqnarray}
Note that the above calculation is nothing but that for deriving the trace anomaly \cite{Chanowitz}.

Next, let us switch on the U(1) gauge fields and calculate the trace anomaly coming from this sector.
The calculation proceeds in a perfectly similar manner, so let us comment on only the essential point. 
The point is that the trace anomaly from this sector is proportional to the square of the mass of the gauge field, i.e.
$\frac{1}{\sqrt{- \hat g}} < \partial_\mu J^\mu > = c g_{BL}^2 M_B^2 M_p^2$ with $c$ being some 
constant of order 1.

Now let us mention the relation between the one-loop results obtained in this section and the conditions  (\ref{Condition})
obtained in the classical analysis. The total trace anomaly takes the form 
\begin{eqnarray}
\frac{1}{\sqrt{- \hat g}} < \partial_\mu J^\mu >
= \frac{24}{\pi^2} \lambda_H  m_h^2 M_p^2 + c g_{BL}^2 M_B^2 M_p^2.
\label{Trace anomaly 2}
\end{eqnarray}
Substituting Eq.  (\ref{Parameters}) and the definition $M_B = \frac{2}{\sqrt{\xi}} g_{BL} M_p$ into this relation
leads to 
\begin{eqnarray}
\frac{1}{\sqrt{- \hat g}} < \partial_\mu J^\mu >
= \frac{48}{\pi^2} \lambda_H  \frac{1}{\xi} |\lambda_{H \Phi}|  M_p^4 + 4 c \frac{1}{\xi} g_{BL}^4 M_p^4.
\label{Trace anomaly 3}
\end{eqnarray}
According to a recent study of the trace anomaly in the non-renormalizable theories \cite{Armillis}, this trace
anomaly must be very tiny or zero. Assuming each term in the RHS of (\ref{Trace anomaly 3}) to be independent of each
other, we get the relations 
\begin{eqnarray}
\lambda_H \frac{|\lambda_{H \Phi}|}{\xi} \approx 0, \quad \frac{1}{\xi} g_{BL}^4 \approx 0.
\label{Condition 2}
\end{eqnarray}
In particular, note that the former relation is roughly consistent with  (\ref{Condition}), which means that
these conditions remain unchanged even if radiative corrections are included in our model. As a final remark,
let us comment on higher-order quantum effects more than one-loop. It is easy to see that in higher-loop
amplitudes, the dilaton becomes massive and there is no nice mechanism to prevent the mass from taking 
the Planck mass. As a result, the current conservation is modified as
\begin{eqnarray}
\frac{1}{\sqrt{- \hat g}} \partial_\mu J^\mu = \frac{1}{\zeta} (D-4) V^{(0)} (\hat H) 
+ \frac{1}{\zeta} m^2 \sigma + \cdots,
\label{Total-Trace anomaly}
\end{eqnarray}
where $m^2$ is the mass of the dilaton obtained by radiative corrections and $\cdots$ denotes the other 
higher-order contributions. The second term is linear in the dilaton field, so it does not make any contribution
to the trace anomaly when we take the expectation value. From this consideration, our conclusion in this section
that the conditions  (\ref{Condition}) remain unchanged in the one-loop level, might be true even in the 
higher-order levels.

\section{Discussion}

In this article, we have considered a coupling of dilaton gravity to a classically scale-invariant B-L
extension of standard model and seen that at the tree level, spontaneous symmetry breaking of the U(1) B-L gauge symmetry
occurs and the corresponding gauge field acquires a mass as a result of spontaneous symmetry breakdown 
of scale invariance. Although we have taken account of a specific model, it is obvious to apply our idea
to any model of the standard model extensions with classical scale invariance. In our approach, we implicitly assume
that there is no new physics between the electroweak and Planck scales, and in a sense the electroweak scale is
determined by Planck physics. Then, it is physically reasonable to incorporate the gravity sector into the action.  

Our analysis in this article is confined to the classical and one-loop analyses. Even if the details of the full quantum-mechanical analysis 
of the present theory will be reported in a separate publication, some comments on them in advance might deserve a particular
attention.  If the classical scale invariance were broken at the higher-order loop effects, the dilaton $\sigma$ not only
could become massive but also start to interact with the other fields. As a bonus, the dilaton could then have the right of becoming a candidate 
of dark matter if this particle is sufficiently "cold" and stable. The other interesting aspect of the quantum analysis is that 
the renormalization group has the contribution from gravity sector in addition to that from the standard model extensions
\begin{eqnarray}
\mu \frac{d \lambda_i}{d \mu} = \beta_i^{SME} + \beta_i^{GR},
\label{RG}
\end{eqnarray}
where on dimensional grounds the beta function $\beta_i^{GR}$ from the gravity sector takes the form
\begin{eqnarray}
\beta_i^{GR} = \frac{c_i}{8 \pi} \frac{\mu^2}{M_p(\mu)^2} \lambda_i,
\label{Beta}
\end{eqnarray}
with the coefficients $c_i$ depending on the detail of the gravity sector. Thus it is of interest to calculate the coefficients $c_i$
by an explicit calculation to examine the stability bound of the Higgs mass. 

As another interesting study for an application of our idea, we could list up the Higgs inflation \cite{Bezrukov}. Let us note that our Lagrangian 
density is not most general in the sense that one can add one more renormalizable and scale-invariant term, which is
$\sqrt{-g} H^\dagger H R$. This term plays a critical role in the scenario of the Higgs inflation, but needs an additional field in order to
avoid violation of unitarity. Since our model includes a complex single scalar $\Phi$, there might be a possibility of realizing
the Higgs inflation without unitarity violation. This issue will be also reported in the future.

\begin{flushleft}
{\bf Acknowledgements}
\end{flushleft}

This work is supported in part by the Grant-in-Aid for Scientific 
Research (C) No. 22540287 from the Japan Ministry of Education, Culture, 
Sports, Science and Technology.


\end{document}